\documentclass{article}
\usepackage{fullpage}
\AtBeginDocument{%
  \providecommand\BibTeX{{%
    \normalfont B\kern-0.5em{\scshape i\kern-0.25em b}\kern-0.8em\TeX}}}

\usepackage{url}
\usepackage{booktabs}
\usepackage{xcolor}
\usepackage{graphicx}
\usepackage{amssymb}
\usepackage{amsmath}
\usepackage{graphicx}
\usepackage{latexsym}
\usepackage{epstopdf}
\usepackage{multicol}
\usepackage{multirow}
\usepackage{fancyvrb}
\usepackage{subcaption}
\usepackage[skip=0.5pt]{caption}
\usepackage{tikz}
\usepackage{paralist}
\usepackage[vlined,ruled,commentsnumbered]{algorithm2e}
\let\oldnl\nl
\newcommand{\nonl}{\renewcommand{\nl}{\let\nl\oldnl}}%
\usepackage{tikz}
\usetikzlibrary{trees,arrows.meta,chains,fit,shapes,calc}
\usepackage{xcolor,listings,colortbl}
\usepackage{textcomp}
\usepackage{forest}
\usepackage{mathtools}
\usepackage{enumitem}
\usepackage{tikz-qtree}
\usepackage{pgfplots}
\usepackage{tabularx}
\usepackage{fancyvrb}
\usepackage{array}
\usepackage{dsfont}
\usepackage{balance}  
\usepackage{hyperref}
\usepackage{amsthm}

\makeatletter
\def\@copyrightspace{\relax}
\makeatother

\tikzset{cross/.style={cross out, draw, 
         minimum size=2*(#1-\pgflinewidth), 
         inner sep=0pt, outer sep=0pt}}
\definecolor{myblue}{RGB}{80,80,160}
\definecolor{mygreen}{RGB}{80,160,80}

\makeatletter
\newcommand{\printfnsymbol}[1]{%
  \textsuperscript{\@fnsymbol{#1}}%
}
\makeatother

\SetCommentSty{mycommfont}

\newcommand{\reminderb}[1]{\textcolor{red}{[[[#1]]]}}

\newcommand{\reminderd}[1]{\textcolor{orange}{[[[#1]]]}}

\newcommand{\cut}[1]{{}}

\newcommand{\amir}[1]{\reminderb{\bf (Amir)~#1}{\typeout{#1}}}

\newcommand{\harsh}[1]{\reminderd{\bf (Harsh)~#1}{\typeout{#1}}}

\definecolor{Gray}{gray}{0.85}
\newcolumntype{a}{>{\columncolor{Gray}}c}

\newcommand{\paratitle}[1]{\noindent{\bf #1}}

\newcommand{\set}[1]{\boldsymbol{#1}}
\newcommand{\ds}{\ensuremath{\delta^2}}
\newcommand{\is}{\ensuremath{\iota^2}}

\newtheorem{theorem}{Theorem}[section]

\newtheorem{definition}[theorem]{Definition}

\let\OLDthebibliography\thebibliography
\renewcommand\thebibliography[1]{
  \OLDthebibliography{#1}
  \setlength{\parskip}{0pt}
  \setlength{\itemsep}{0pt plus 0.3ex}
}

\definecolor{keywords}{rgb}{0,0,0.7}

\title{Heterogeneous Treatment Effects in Social Networks}

\author{ Amir Gilad\thanks{Both authors contributed equally to this research.} \thanks{Duke University} , Harsh Parikh\printfnsymbol{1}\printfnsymbol{2},  Sudeepa Roy\printfnsymbol{2}, Babak Salimi\thanks{University of California, San Diego}}

\begin{document}
\maketitle

\begin{abstract} 
We study {\em treatment effect modifiers} for causal analysis in a social network, where neighbors' characteristics or network structure may affect the outcome of a unit, and the goal is to identify sub-populations with varying treatment effects using such network properties. We propose a novel framework for this purpose that facilitates data-driven decision making by testing hypotheses about complex effect modifiers in terms of network features or network patterns  (e.g., characteristics of neighbors of a unit or belonging to a triangle), and by identifying sub-populations for which a treatment is likely to be effective or harmful. We describe a hypothesis testing approach that accounts for a unit's covariates, their neighbors' covariates, and patterns in the social network, and devise an algorithm incorporating ideas from causal inference, hypothesis testing, and graph theory to verify a hypothesized effect modifier. In addition, we develop a novel algorithm for discovery of network patterns that are potential effect modifiers. We perform  extensive experimental evaluations with a real development economics dataset about the treatment effect of belonging to a financial support network called self-help groups on risk tolerance, and also with a synthetic dataset with known ground truths simulating a vaccine efficacy trial, to evaluate our framework and algorithms.


\cut{
Causal inference methods aims at estimating the effect of a treatment or an intervention on an outcome of interest. Literature has shown that the response to treatment might be different for different individuals \cite{}, i.e. treatment effects are heterogeneous across the population.
Effect modifiers are variables that are associated with the heterogeneity in the treatment effects. For example, elderly patients are more sensitive to propofol \cite{}. The knowledge pf treatment effect modifiers are often helpful because it helps decision makers regulate treatment i.e. where and how to intervene. Previous works in literature have focused on detecting heterogeneity
.....
As an example, knowledge of effect modifiers can help epidemiologists design strategy for vaccination.  
\harsh{stopped here!}
\amir{continue}

However, this assumption does not always capture the social network model that expresses interactions between subjects. 
In particular, when testing for treatment effect modifiers, we can no longer consider just the covariates of each subject, but also have to account for their social environment. 
In this work, we provide a test designed to reveal such properties of the social network as effect modifiers. We devise a causal model for social networks and provide a framework for testing effect modifiers that includes a novel test and algorithms for estimating effect modifiers. 
We provide an experimental study with synthetic and real data that shows the capability of our solution to detect these attributes both in the presence of noise and in with a varying size of network. 
}   
\end{abstract}

\vspace{-3mm}\section{Introduction}\label{sec:intro}
Causal inference is at the heart of empirical research and principled decision-making in natural and social sciences, and is practically indispensable in epidemiology \cite{robins2000marginal}, clinical research \cite{clinicaltrials2007}, 
public policy \cite{manski2013public}, economics \cite{banerjee2011poor}, and other domains \cite{angrist1996identification,angrist2008mostly}. In causal inference, we go beyond establishing correlation or prediction, and are interested in making causal statements using the concepts of \emph{counterfactuals} and \emph{interventions} about a \emph{treatment} (e.g., administering a vaccine) on some \emph{outcome} of interest (e.g., not being infected by the targeted disease).



Recently, there has been a surge of interest in drawing causal inference from social networks (see e.g.,~\cite{vanderweele2013social, ogburn2017causal,awan2020almost, DBLP:conf/uai/ShermanS19,
salimi2019capuchin, ogburn2020causal, kao2017causal, fafchamps2015causal, zheleva2021causal}. In such settings, an individual’s behavior, treatment assignment or outcome could influence their social contacts’ behavior, treatment or outcome. In this paper, we address the problem of {\em detecting treatment effect modifiers~\cite{pearl2017detecting} for causal analysis in a social network}. In a causal study, different units may exhibit different levels of response to the applied treatment based on their characteristics, and the {\em effect heterogeneity} problem aims to infer this degree
of causal effects on different units. 
In healthcare, public policy and education research, identifying treatment effect modifiers can allow optimizing precision-decision-making by reducing the cost and negative side-effects, and maximizing the intended positive effect of treatment.
As an example, one of the key objectives in clinical research is not only to discover effects of 
medical treatments at the entire population-level, but also to ascertain 
treatment effects on 
different sub-populations (e.g., whether a new vaccine is beneficial or harmful based on age or previous health conditions of individuals).

While different hypothesis tests for the heterogeneity of a population in a study have been proposed in the literature \cite{viechtbauer2007hypothesis, athey2016recursive}, along with a large body of work on testing statistical hypotheses (e.g., \cite{snyder1978hypothesis,xia2017hypothesis,LAROSA201581,shaffer1995multiple,list2019multiple}), such tests 
do not take into account the context of the social network like possible effects of the neighbors' covariates or the network structure on the outcome. On the other hand, the existing methods in causal inference from social networks primarily focused on the average treatment effects. Indeed, reasoning about effect modification and heterogeneity in social networks
poses new conceptual and computational challenges.

\subsection*{Our Contributions}
We propose a novel framework 
for the estimation of treatment effect heterogeneity and detecting effect modifiers, tailored to causal inference in social networks. 
Our framework facilitates the following aspects to assist in discovering effect modifiers in social networks: (1) mining network patterns that are suspected effect modifiers  through a novel criterion and algorithm, (2) testing for existence or lack of existence for various kinds of heterogeneous treatment effects using summarized properties of network features or neighborhood structures, (3) identification of sub-populations for which a treatment is effective, neutral, or harmful, (4) generalization of causal effect estimates and causal conclusions obtained from a study sample to a target population based on relative distributions of different sub-populations.
Our technical contributions in the paper are as follows.

\noindent
\textbf{Model, framework, and guarantees:} 
We define a causal model for social networks that captures complex effect modifiers that may arise because of complex social interaction between the units, i.e., effect modifiers associated with the neighbors' covariates on the outcome and the network structure associated with each unit. 
We then propose a novel hypothesis testing approach that is analogous to the $I^2$ test for heterogeneity in meta-analysis \cite{higgins2003measuring}. Our testing framework allows for neighbors' covariates and patterns in the social network as effect modifiers.
Since the number of neighbors of each unit in the network may be different, we use the concept of covariate summary proposed in 
\cite{DSalimiPKGRS20}. Given a covariate of a unit in the social network, we aggregate this covariate across all neighbors into a single summarized number (e.g., the percentage of neighbors who work as Farm Labour). 
We then formally prove that the consistency of the framework. 
Next, we define the problem of mining network patterns that are suspected effect modifiers. We describe it formally as an optimization problem that aims to maximize the difference between the treatment effect of nodes whose neighborhood contains the pattern and nodes whose neighborhood does not contain them. We further ensure that the pattern is prevalent in the network to reduce the effect of outliers. 




\paratitle{Algorithm for testing:} 
We devise an algorithm for testing effect modifiers. 
The algorithm incorporates ideas from causal inference and hypothesis testing. In particular, our solution uses existing causal inference techniques from the literature to compute the treatment effect for each unit (conditional treatment effect) and the value of the hypothesized effect modifier for each unit. 
Once these are computed, it estimates the influence of the covariate on the treatment effect. 

\paratitle{Experimental study:} We provide an extensive experimental evaluation using the real social network described in our example \cite{jackson2012social,banerjee2014gossip,microfinancedata} as well as synthetic data with known ground truth, simulating a vaccine efficacy trial. 
Through the synthetic data, we examine the effect of the number of units and the noise in the potential outcome on our test. 
Our results indicate that as the number of units increases, 
the true effect modifiers become more evident and that our solution becomes relatively insensitive to high levels of noise. 
Through the real data, we demonstrate a use-case, showing our system's performance and its ability to find effect modifiers in a large-scale network ($\sim$17K units). Some of the effect modifiers have not been previously discovered, to our knowledge, and are based on summarized neighbors' covariates and some are network patterns that have. Our code will be public.
\cut{
\paratitle{Related work}\label{sec:related}
There is rich literature on statistical tests for hypothesis testing 
\cite{snyder1978hypothesis,shaffer1995multiple,list2019multiple,xia2017hypothesis,LAROSA201581}, heterogeneity detection \cite{hardy1998detecting,crump2008nonparametric,green2012modeling,taddy2016nonparametric}, effect modification \cite{katsouyanni2001confounding,greenland1989ecological,knol2012recommendations,vanderweele2009distinction,vanderweele2007four,hernan2010causal,blot1979synergism,rothman1980concepts,saracci1980interaction}, and causal inference in the presence of interference \cite{graham2010measuring,halloran2012causal,halloran1995causal,vanderweele2011bounding,aronow2017,shalizi2011homophily,ogburn2017causal}. 
This is the first paper, to our knowledge, that provides a test for effect modifiers that is tailored for the unique aspects of social networks, namely, detecting effect modifiers that stem from neighbors' covariates and from the structure of the network itslef. 
}

\section{Causal Inference and
Networks}\label{sec:background}
In this section, we
discuss relevant concepts about causal inference and social networks. As a convention, capital letters are used to denote random variables, hats $\widehat{\cdot}$ are used for estimates, and \textbf{bold} capital letters are used to denote sets or vectors.
\subsection{Setup}\label{sec:setup}
We consider a dataset of $n$ units, connected with each other in a social network $\set{G}(\set{V}, \set{E})$, e.g., 
a network of villagers connected with each other in a friendship or relationship network. $\set{G}$ contains a set of units, $\set{V} = \{v_1,\dots,v_n\}$, and a set $\set{E}$ of (undirected) ties between the units, where $E_{i,j}$ denotes the edge between nodes $v_i$ and $v_j$. 
We assume that $E_{i,i}=1$ for all $v_i\in\set{V}$. For 
each $v_i \in \set{V}$, we have information about their pre-treatment covariates $\set{X}_i$, observed post-treatment outcome $Y_i$, and choice of treatment $T_i$. We assume the treatment variable to be binary, however, our framework generalizes to n-ary treatments. Further, we define $Y_i(t)$ to be the potential outcome for treatment choice $t \in \{0, 1\}$ \cite{rubin1970thesis,pearl2000book}. Thus, under the no-interference assumption, the observed outcome can be represented in terms of potential outcome and treatment choice as: $Y_i = T_i Y_i(1) + (1-T_i) Y_i(0)$.

For $v_i \in\set{V}$, the \emph{ego-centric network} is a sub-network $\set{G}_i = (\set{V}_{\set{G}_i}, \set{E}_{\set{G}_i})$ such that $\set{V}_{\set{G}_i} = \{ v_j ~|~ E_{i,j}=1 \} $ (hence $v_i \in \set{V}_{\set{G}_i}$) and $\set{E}_{\set{G}_i} = \{ E_{j,k} ~|~ v_j,v_k \in \set{V}_{\set{G}_i}\}$.

\subsection{Probabilistic Causal Model for Networks} 
\label{sec:pcmsn}
We use probabilistic causal models and structural equations \cite{pearl2000book}
to define a data generative model for social networks that enable us to reason about the effects of interventions (our model is similar to prior work, e.g., \cite{ogburn2017causal}). 

\paratitle{Causal models for social networks.}
A probabilistic causal model for a social network $\set{G}(\set{V}, \set{E})$ with observed ties $\set{E}$ is a tuple $\set{M} = (\set{\epsilon}, \set{Z}, 
Pr_{\set{\epsilon}}, \mathbf{\Phi})$, 
where 

\vspace{-2mm}
\begin{itemize}
\itemsep0em
    \item $\set{\epsilon}=\{\epsilon^X_i\}_i\cup\{\epsilon^T_i\}_i\cup\{\epsilon^Y_i\}_i\cup\{\epsilon^E_{i,j}\}_{i,j}$ is a set of unobserved {\em exogenous} variables corresponding to ${\set X}_i$, $T_i$, $Y_i$, and $E_{i,j}$ distributed according to $Pr_{\set{\epsilon}}$,
    \item $\set{Z} = (\set{X},\set{Y},\set{T})$ is a set of {\em observed} ({\em endogenous}) variables,
    \item $\mathbf{\Phi}$ 
    is a set of {\em structural equations} described below. 
\end{itemize}
\vspace{-2mm}

\paratitle{Structural equations and causal dependency.} 
We define $\mathbf{\Phi} = \{\phi_X,\phi_T,\phi_E,\phi_Y\} $ as the set of {\em structural equations} that describes generative process and causal dependence of the observed variables. (1) We assume that the pre-treatment covariates $\set{X}_i$ are only functions of exogenous variables $\{\epsilon^X_i\}$ of the form $\set{X}_i = \phi_X(\epsilon^X_i)$. 
(2) The network ties 
are a function of units' covariates and pair-wise exogenous variables $\epsilon^E_{i,j}$ for units $v_i$ and $v_j$ of the form $E_{i,j} = \phi_E(\set{X}_i,\set{X}_j,\epsilon^E_{i,j})$ and $E_{i,i}=1$ by convention. (3) A unit's 
treatment $T_i$ is determined by 
$\phi_T(\set{X}_i,\{\set{X}_j\}_{j\in \set{V}_{\set{G}_i}\setminus\{v_i\}}, \set{E}_{\set{G}_i},\epsilon^T_i)$. (4) Finally, the post-treatment outcomes $Y_i$ is determined by 
$\phi_Y\left(\set{X}_i,\{\set{X}_j\}_{v_j\in \set{V}_{\set{G}_i}\setminus\{v_i\}}, \set{E}_{\set{G}_i}, T_i \right) + \epsilon^Y_i$. Thus, the covariates of 
$v_i$ itself, their neighbors' covariates, and edges in its ego-centric network (i.e., $\set{X}_i,\{\set{X}_j\}_{v_j\in \set{V}_{\set{G}_i}\setminus\{v_i\}}, \set{E}_{\set{G}_i}$) are potential confounders that affect both the treatment $T_i$ and post-treatment outcome $Y_i$ of $v_i$.
 
We make the following common assumptions 
in causal inference and social networks literature \cite{ogburn2017causal, rosenbaum1984reducing,fullversion}:
 
  
 \paratitle{(1) Distributional assumptions.} 
 We assume that the 
 unobserved exogenous variables are independent, i.e.,
 for all $v_i, v_j$, (A.1) $\epsilon^X_i \perp \epsilon^X_j$, (A.2) $\epsilon^T_i \perp \epsilon^T_j$, (A.3) $\epsilon^Y_i \perp \epsilon^Y_j$, and (A.4) for all $v_i$, $E[\epsilon^Y_i]=0$. 

\cut{
\begin{enumerate}[label=A.\arabic*]
    \item $\epsilon^X_i \perp \epsilon^X_j$ for all units $i$ and $j$
    \item $\epsilon^T_i \perp \epsilon^T_j$ for all units $i$ and $j$
    \item $\epsilon^Y_i \perp \epsilon^Y_j$ for all units $i$ and $j$
    \item $E[\epsilon^Y_i]=0$ for all $i$
    \item $\epsilon^E_{i,j} \perp \epsilon^E_{j,k} \perp \epsilon^E_{k,l} \perp \epsilon^E_{i,k}$ for all units $i$,$j$,$k$ and $l$
\end{enumerate}
}

 \paratitle{(2) Summarizability assumption.}  We assume the existence of 
 functions $s_X$ and $s_E$ that summarize covariates of neighbors and ego-centric network: if $\omega_i = s_X(\{\set{X}_j\}_{v_j\in \set{V}_{\set{G}_i} \setminus\{v_i\}})$ and $\eta_i = s_E(\set{E}_{\set{G}_i})$ then 
\begin{small}
\begin{equation*}\label{eq:assumption_summary}
    \phi_Y\left(\set{X}_i,\{\set{X}_j\}_{v_j\in \set{V}_{\set{G}_i}\setminus\{v_i\}}, \set{E}_{\set{G}_i}, T_i \right) = \phi_Y\left(\set{X}_i,w_i, \eta_i, T_i \right)
\end{equation*}
\end{small}
 This is assumption is useful because different units might have different number of neighbors.

 \paratitle{(3) Positivity assumption.} Finally, we assume the that propensity of any unit $v_i$'s treatment 
 $T_i$ is bounded away from 0 and 1, i.e,
 {\footnotesize
 \begin{equation*}\label{eq:assumption_positivity}
     0 < P(T_i = 1 | \omega_i, \eta_i, \set{X}_i) < 1 \;\; \forall v_i \in \set{V}
 \end{equation*}
 }
\paratitle{Network patterns.} A network pattern $\Delta$ is a collection of nodes and (undirected) edges connected to the nodes. For instance, there are four possible patterns with three nodes (three isolated nodes, an edge and a node, a path of length 2, and a triangle). Let ${\cal P}$ denote a set of network patterns (we consider only patterns with small set of nodes for efficiency and interpretability). 
For a pattern $\Delta \in {\cal P}$ and the ego-centric network $\set{G}_i$ of a unit $v_i$, we say $\Delta \in \set{G}_i$ if $\Delta$ is {\em isomorphic} to a subgraph in $\set{G}_i$.

\paratitle{Treatment effect.} 
The effect of a binary treatment $T_i$ on an outcome $Y_i$ is measured by comparing the potential outcomes $Y_i(1)$ and $Y_i(0)$. Formally, treatment effect for unit $v_i$ is defined as $\tau_i = Y_i(1) - Y_i(0)$. However, for any given unit, we only observe one of the potential outcomes given the treatment choice, and 
it is impossible to know the true treatment effect for any unit. 
%
Given a set of endogenous variables and network patterns $\set{W} \subseteq \set{X}\cup {\cal P}$, following are the estimands of interest:
(1) Average Treatment Effect (ATE): $\overline{\tau} = \mathbb{E}[Y_i(1) - Y_i(0)]$, (2) average Treatment Effect on Treated (ATT): $\overline{\tau}^{(1)} = \mathbb{E}[Y_i(1) - Y_i(0) | T_i = 1]$, (3) Conditional ATE (CATE): $\tau(\set{w}) = \mathbb{E}[Y_i(1) - Y_i(0) | \set{W}_i = \set{w}]$, (4) Conditional ATT (CATT): $\tau^{(1)}(\set{w}) = \mathbb{E}[Y_i(1) - Y_i(0) | \set{W}_i = \set{w}, T_i = 1]$.
Here $\set{w}$ denotes a set of values from the domain of $\set{W}$ (indicator variables for patterns).  We use $\set{W}_i = \set{w}$ to denote $\Delta \in \set{G}_i$ for all patterns $\Delta \in \set{W}$, and $X_i = x$ for all covariates $X \in \set{W}$ 
with 
values $x \in {\bf w}$.
\subsection{Treatment Effect Modifiers}
Given $\set{W} \subseteq \set{X}\cup {\cal P}$, if there exist two values $\set{w}$ and $\set{w}'$ in the domain of $\set{W}$ such that $\tau(\set{w})\neq \tau(\set{w}')$ or $\tau^{(1)}(\set{w})\neq \tau^{(1)}(\set{w}')$, i.e., the CATE or CATT varies according to the two sub-populations with $\set{W}=\set{w}$ and $\set{W}=\set{w}'$, then the treatment effect is {\em heterogeneous} and the variables $\set{W}$ are called {\em treatment effect modifiers} (or simply \emph{effect modifiers}) \cite{hernan2010causal,vanderweele2007four, athey2016recursive}. 

\cut{
In healthcare, public policy and education research, identifying treatment effect modifiers can allow optimizing precision-decision-making by reducing the cost and negative side-effects, and maximizing the intended positive effect of treatment. For instance, if a doctor knows that certain drug is more helpful to elderly patients but is quite harmful to younger patients, then the doctor can design highly efficacious treatment regimes benefiting as many patients as possible. 
}

\paratitle{Problem statement.} In this paper, we are interested in estimating treatment effects for a given social network data and identifying if the treatment effect is heterogeneous across different levels of unit's covariates, their social network neighbors' covariates, and the structure of their ego-centric graph. Specifically, our 
goal is to find a set of variables that are effect modifiers, i.e., 
we want to find variable(s) $\set{W}$ such that:

\begin{small}
\begin{equation*}
    \exists \set{w} \in Dom(\set{W}) \textrm{ such that } \overline{\tau} \neq \tau(\set{w}), \text{ or, } \overline{\tau^{(1)}} \neq \tau^{(1)}(\set{w}) 
\end{equation*}
\end{small}
\paratitle{Challenge with finite sample.} Given a finite sample from the population, 
estimated ATE ($\widehat{\overline{\tau}}$) and CATEs ($\widehat{\tau}(\set{w})$) 
are likely to be always unequal.
However, it is possible that $\widehat{\tau}(\set{w})$ might converge to $\widehat{\overline{\tau}}$ as $n\rightarrow\infty$. Thus, we need to infer if the observed difference between $\widehat{\overline{\tau}}$ and $\widehat{\tau}(\set{w})$ is statistically significant. Thus, we focus on developing an inference framework to test for effect modifiers (Section~\ref{sec:model}).

\paratitle{Challenge with social network data.} In our setup, we allow the ego-centric network to be a potential confounder as well as effect modifier. Hence, it is possible that the social network neighbors' covariates can also be effect modifier, e.g., a unit $v_i$'s friends having or not having bank accounts can affect effectiveness of unit's participation in a self-help group. Similarly, existence of certain network sub-structures (such as a triangular relationship) in ego-centric network can also be potential effect modifier. For instance, participation in self-help group might be more beneficial to individuals who are part of a clique of size 3 or more compared to units who are not. Mining an effect modifying pattern ($\Delta$) is non-trivial as search over all possible patterns in an ego-centric graph is practically infeasible given the exponentially large number of potential patterns. Thus, we develop a scalable pattern mining algorithm which can finding interesting patterns $\Delta$ that are likely to be effect modifiers (Section~\ref{sec:single}).

\section{Framework}\label{sec:model}
In this section, we first describe our testing framework for effect heterogeneity by formally introducing the concept of \emph{hypothesized effect modifiers} and the testing criterion (Section~\ref{sec:criteria_test}) and provide theoretical guarantees (Section~\ref{sec:theory}). Then we delineate the estimation and testing procedure used to operationalize the framework 
along with our algorithm (Section~\ref{sec:single}).

\subsection{Criterion for testing}\label{sec:criteria_test}

\cut{
\begin{definition}[Hypothesized effect modifiers]
Given a social network $\set{G} = (\set{V},\set{E},\set{Z})$, 
a hypothesized effect modifier 
is a random variable 
which can be one of (1) the covariates of the unit $i$, $\set{X}_i$, (2) summarized covariate for unit $i$’s neighbors $s_X(\set{X}_{\set{V}_{\set{G}_i}})$, or (3) indicator for a network pattern $\Delta$ in $\set{G}_i$.
\end{definition}

Intuitively, we are interested in checking whether a hypothesized covariate $W$ is an effect modifier or not. If $W$ is an effect modifier then $\Omega_W = \mathbb{E}_W[(\tau(W) - \bar{\tau})^2] >0$, otherwise $\Omega_W=0$.
}

\paratitle{Hypothesized effect modifiers.}
Given a social network $\set{G} = (\set{V},\set{E})$ and its causal model $M$ with observed variables $\set{Z}  = (\set{X}, \set{Y}, \set{T})$, 
a hypothesized effect modifier 
is a set of 
variables $\set{W} \subseteq (\set{X} \cup {\cal P})$ such that 

\begin{itemize}
    \item each variable in $\set{W}$ is one of (1) a covariate $X \in \set{X}_i$ of a unit $v_i$ itself, (2) a summarized covariate $X$ for unit $v_i$’s neighbors $s_X(X_{\set{V}_{\set{G}_i \setminus \{v_i\}}})$, or (3) 
    a network pattern $\Delta$ such that $\Delta \in \set{G}_i$, and 
\item 
either $\Omega_W = \mathbb{E}_W[(\tau(W) - \bar{\tau})^2] >0$ (for CATE), or $\Omega^{(1)}_W = \mathbb{E}_W[(\tau^{(1)}(W) - \bar{\tau^{(1)}})^2] >0$ (for CATT).
\end{itemize}

 
If $W$ is not an effect modifier, then  $\Omega_W = 0$.
Testing if $\Omega_W$ is non-zero is analogous to the problem of testing for heterogeneity in meta-analysis. Thus, we adapt and build on the $I^2$ measure proposed by \cite{higgins2002quantifying,higgins2003measuring} as a test statistic of interest. $I^2$ is a composite measure defined in terms of Cochran's $Q$ measure of heterogeneity and the statistical degree of freedom (df): $I^2 = \frac{Q - \text{df}}{Q}$. 
If the computed value of $I^2$ is negative, it is rounded up to zero. 
Intuitively, $I^2$ describes the proportion of total variation across studies that can be attributed to heterogeneity. 
%
%
We adapt these measures to our framework of causal inference for social network to test for effect modifiers, and handle continuous covariates. 
For the given social network $\set{G} = (\set{V},\set{E})$ and a hypothesized covariate or pattern $W$, we measure the heterogeneity due to $W$ 
using $\ds$ which is analogous to the normalized version of Cochran's Q 
\cite{higgins2002quantifying,higgins2003measuring,cochran1954combination}.

{\footnotesize
\begin{align*}
    \ds_W\ = \int_{w} \frac{\left(  \tau(w) - \bar{\tau} \right)^2}{\nu^2(w)} p_{W}(w) dw
\end{align*} 
}

where $p_{W}$ is the probability density function for covariate $W$ and $\nu^2(w) = \mathbb{E}[ (Y_{i} (1) - Y_{i} (0))^2 | W = w ] - \tau^2(w)$.
Using $\ds_W$ we compute our primary test statistic $\is_W$ (which is analogous to $I^2$ described in \cite{higgins2002quantifying}) as:
{\footnotesize
 \begin{equation}
     \is_W = \frac{\ds_W - 1}{\ds_W}
 \end{equation}
 }
 The null hypothesis for the test for treatment effect modifier for the hypothesized covariate $W$ is the absence of heterogeneity across the different strata of $W$.
The test rejects the null hypothesis if the estimated $\widehat{\is}_W$ using the observed data is 
larger than or equal to a predetermined threshold $I_0$ (in our experiments, we observed that setting $I_0 = 0$ has desired performance).

\subsection{Theoretical Guarantees}\label{sec:theory}

First, we prove (proofs in the supplementary material) that the conditional average treatment effect of interest are identifiable in terms of observables -- $\set{X},\set{Y},\set{T}$ and $\set{E}$. This is important to ensure that we design a method that can estimate the treatment effect using the finite data.

\begin{theorem}[Identification of causal effects]\label{thm:identify}
The causal average treatment effect $\tau(\set{X}_i, \{\set{X}_j\}_{v_j\in \set{V}_{\set{G}_i}\setminus\{v_i\}}, \set{E}_{\set{G}_i}) = E[Y_i(1) - Y_i(0) | \set{X}_i, \{\set{X}_j\}_{v_j\in \set{V}_{\set{G}_i}\setminus\{v_i\}}, \set{E}_{\set{G}_i}]$ is identified as a function of observed variables $\set{X},\set{Y},\set{T}$ and $\set{E}$.
\end{theorem}

Now, we show that if we consistently estimate the treatment effects then the test is consistent under null, 
i.e., the type 1 error of our test 
diminishes as the size of the data gets larger. 

\begin{theorem}[Consistency under null]\label{thm:consistency}
Given a consistent estimator of conditional average treatment effects and average treatment effects, and $I_0=0$, the test statistic $\widehat{\is}_W$ is asymptotically consistent under null i.e. if feature W is not an effect modifier then $\widehat{\is}_W \to 0$ as $n \to \infty$.
\end{theorem}

\cut{ Next, we delineate our estimation and testing procedure in Section~\ref{sec:single}. Further, we discuss the identification of treatment effects and consistency of the test in Section~\ref{sec:theory}.
}

\subsection{Estimation and Testing Procedure}\label{sec:single}
In this section we discuss each component of our estimation and testing procedure, and show how we combine them to infer whether a hypothesized effect modifier is an effect modifier using $\is$ described in the previous section. 
We restrict our discussions to single network patterns $\set{W} = \{\Delta\}$ in this section for simplicity (extensions to covariates and sets are discussed in the supplementary material.)
Our approach has four steps: (1) mining a network structure that can be a potential effect modifier, (2) CATE estimation, (3) CATE smoothing and variance estimation, and (4) hypotheses testing.

\paratitle{Pattern Mining.} 
For patterns $\Delta$ that are potential effect modifiers, the average treatment effect for units $v_i$ that have $\Delta \in \set{G}_i$ 
is different from the average treatment effects of the other units. This can be translated to the following optimization problem:

\begin{equation*}
    \textrm{arg max}_{\Delta} \left| \mathbb{E}\left[ \tau_i | \Delta \in \set{G}_i \right] - \mathbb{E}\left[ \tau_i | \Delta \notin \set{G}_i \right] \right|.
\end{equation*}

Intuitively, this objective function is inspired by the definition of effect modifiers in the classical causal inference literature \cite{rubin2005causal}, and aims to find a pattern for which the heterogeneity in CATE is maximized. However, as  mentioned earlier, we do not know the treatment effect for each unit and estimating the 
CATE for each potential $\Delta$ is computationally expensive. 

We can note from the structural equations discussed in Section~\ref{sec:pcmsn} that a pattern can be a treatment effect modifier if the outcome is heterogeneous for $\Delta \in \set{G}_i$ and $\Delta \notin \set{G}_i$. It is important to note, however, that heterogeneity of outcome $Y_i$ does not imply heterogeneity of treatment effect $\tau_i$. Thus, modifying the optimization object as follows can assist mining of interesting pattern(s) which are later tested for treatment-effect heterogeneity after estimation:

\begin{equation*}
    \textrm{arg max}_{\Delta} \sum_t \left| \mathbb{E}\left[ Y_i | \Delta \in \set{G}_i, T_i=t \right] - \mathbb{E}\left[ Y_i | \Delta \notin \set{G}_i, T_i=t  \right] \right|
\end{equation*}

Thus, for a given a finite sample with $n$ units, we want to find a pattern that maximizes the difference between the outcome of nodes that contain the pattern in their ego-centric network and the outcome of those that do not contain the pattern in their ego-centric network for a given treatment arm. Additionally, we want the pattern to be prevalent in the network. In other words, we want to remove the possibility of the pattern being present in the ego-centric networks of a few outlier nodes with exceptionally high outcomes that contain the pattern, while the rest of the nodes with high outcomes do not contain it. 
This is handled by the constraint over the difference between the nodes whose ego-centric network contains the pattern and those whose ego-centric network does not whose contain the pattern. Formally:

\begin{equation*}
    \text{arg max}_{\Delta} \sum_{t\in\{0,1\}}
    \begin{vmatrix}
         \frac{\sum_i \mathds{1}[\Delta \in \set{G}_i] \mathds{1}[T_i=t] Y_i}{\sum_i \mathds{1}[\Delta \in \set{G}_i] \mathds{1}[T_i=t] } - \frac{\sum_i \mathds{1}[\Delta \notin \set{G}_i] \mathds{1}[T_i=t] Y_i}{\sum_i \mathds{1}[\Delta \notin \set{G}_i] \mathds{1}[T_i=t] } 
    \end{vmatrix}
\end{equation*}

\begin{equation*}
\text{such that } \left|\frac{\sum_i \mathds{1}[\Delta \in \set{G}_i]}{n} - \frac{\sum_i \mathds{1}[\Delta \notin \set{G}_i]}{n}  \right| \leq \delta
\end{equation*}

where $0 \leq \delta \leq 1$. The parameter $\delta$ can be chosen according to the desired sensitivity to outliers where $\delta = 1$ means that the pattern $\Delta$ can be found in the ego-centric network of a single node, i.e., there is no sensitivity to outliers, and $\delta = 0$ means that the pattern has to be in exactly half of the ego-centric networks. We employ a greedy approach for pattern mining, detailed in the supplementary material, as the number of patterns is exponential in the graph size.

\paratitle{Causal effect estimation.}
Our causal effect estimation framework respects the structural equations and assumptions discussed in Section~\ref{sec:background}. Recall that based on the structural equations $\phi_T$ and $\phi_Y$,
$\set{X}_i$,$\{\set{X}_j\}_{v_j\in \set{V}_{\set{G}_i}\setminus\{v_i\}}$, $\set{E}_{\set{G}_i}$ are confounders that affects both the treatment choice and post-treatment outcome. Further, based on the summarizability assumption (equation~\ref{eq:assumption_summary}), we can summarize $\{\set{X}_j\}_{v_j\in \set{V}_{\set{G}_i}\setminus\{v_i\}}$ as $\omega_i$ and $\set{E}_{\set{G}_i}$ as $\eta_i$. (In practice, we use $mean$ as $s_X$ and we choose $s_E$ to return the largest eigenvalue of adjacency matrix of ego-centric along with the vector of indicators for existence of mined pattern(s) in the ego centric network. However, one can use different summary function based on the domain knowledge.) Thus, adjusting for $\set{X}$, $\omega$ and $\eta$ is sufficient to estimate the missing potential outcomes and subsequently conditional average treatment effect. We use \textit{non-parametric double machine learning} method with gradient boosting regression to adjust for the above-mentioned variables \cite{chernozhukov2018double}. Non-parametric double machine learning allows consistent estimation of CATEs even when the propensity score or prognostic scores are non-linear. It also protects against model specification.


\paratitle{CATE smoothing and variance estimation.} We project the estimated CATEs on to the space of hypothesized effect modifier(s) $\set{W}$ using standard non-parametric regression approaches such as gradient boosting regression or Bayesian additive regression trees \cite{friedman2001greedy}. This projection is helpful in smoothing the estimates of conditional average treatment effects. We also use the uncertainty quantified (in terms of prediction interval or credible intervals) by the regressor as the estimate of the conditional variance of the treatment effect, $\widehat{\nu}^2(w)$. The experiments in this paper use gradient boosting regression.

\paratitle{Testing the hypotheses.}  
Algorithm \ref{algo:hypothesis_test} describes our approach for hypothesis testing. Given a social network $\set{G}$, a hypothesized effect modifier set $\set{W}$ and a threshold $I_0$, the algorithm first estimates CATEs and ATE. 
Then, it iterates over all the hypothesized effect modifiers $W$ in $\set{W}$. In each iteration, the algorithm projects CATE estimates on to the space of hypothesized effect modifiers in line. 
It then estimates the $\widehat{\ds}_W$ value for the hypothesized effect modifier $W$, and the corresponding $\widehat{\is}_W$ measure. 
For each hypothesized effect modifier $W \in \set{W}$, if $\widehat{\is}_W$ is larger than the threshold $I_0$, the algorithm rejects the null hypothesis, and otherwise it fails to do so. 

\IncMargin{1em}
\begin{algorithm}[t]

	\SetKwFunction{Test hypothesized effect modifiers}{Test hypothesized effect modifier}
	\SetKwInOut{Input}{input}\SetKwInOut{Output}{output}
	\LinesNumbered
	\Input{
	$\set{G}, \set{X}, \set{Y}, \set{T}$, 
	vector of all possible hypothesized effect modifiers $\set{W}$, 
	threshold 
	$I_0$}
    \Output{For each $W \in \set{W}$, True iff $W$ is an effect modifier w.r.t. the threshold $I_0$}
	\BlankLine
    $CATE,ATE \gets \texttt{causalEffectEstimation}(\set{G},\set{X},\set{T},\set{Y})$\;\label{l:causalinference}
     $g \gets project(CATE, \set{W})$\;\label{l:project}
    \ForEach {$W \in \set{W}$} {
        $\hat{\ds}_W \gets \frac{1}{n}
        \sum_{i=1}^n\frac{\left(\widehat{mean}(g(W_{i})) - ATE\right)^2}{\widehat{var}(g(W_{i}))}$\;\label{l:compute_q}
       $\hat{\is}_W \gets max\left(0, \frac{\hat{\ds}_H - 1}{\hat{\ds}_W}\right) $\;\label{l:i_square}
       \If{$\hat{\is}_W > I_0$}
        {\label{l:cond_start}
            \text{Reject Null Hypothesis for }$W$\; 
        }
	   \Else{
	        \text{Fail to Reject Null for }$W$\;
	   }
	   }
	   \label{l:cond_end}
	\caption{Test hypothesized effect modifiers}
	\vspace{-2mm}
	\label{algo:hypothesis_test}
\end{algorithm}
\section{Experiments and Case Study}\label{sec:experiments}

We have conducted both a synthetic data and a real data study to understand the performance of our framework to estimate and test for effect modifiers.
We examine the following questions:
\begin{compactenum}
    \item Can our test detect the true effect modifiers?
    \item What is the effect of increasing the number of units and noise levels on 
    $q$ and $\is$ in Algorithm \ref{algo:hypothesis_test}?
    \item What are the effect modifiers discovered by our algorithm for real data?
\end{compactenum}

\paratitle{Summary of our results:} 
\begin{compactenum}
    \item Synthetic data: Given a large enough sample, Algorithm \ref{algo:hypothesis_test} output $\is > 0$ for the true effect modifiers and $\is = 0$ for covariates that were not effect modifiers, 
    and the computed $\is$ was proportional to the influence of the effect modifier (Table \ref{tbl:synth_data}).
    \item Synthetic data: 
    As the number of units increased, the fluctuation in $\is$ decreased and stabilized for both effect 
    and non effect modifiers (Figure \ref{fig:num_units_a}). 
    \item Synthetic data: 
    Our test is fairly insensitive to relatively large levels of noise; 
    $\is$ values decreased 
    when noise increased as expected (Figure \ref{fig:noise}).
    \item Real data: Our algorithm 
    returned several effect modifiers in the data, some are 
    summarized neighbors' covariates and network patterns (Table \ref{tbl:case_study}).
\end{compactenum}
We give the configuration, implementation, and running time in the supplementary material. 
\subsection{Synthetic Data: Vaccine Efficacy Trial}\label{sec:synthetic}
We analyze our framework's performance using synthetic data for which we know the underlying ground truth treatment effects and effect modifiers. For this experiment, we generate a social network for a simulation of a vaccine efficacy trial.
Specifically, we are interested in studying the efficacy of vaccine, i.e., chances of infection of an individual if they are vaccinated.
The true effect modifiers in the data are (according to their relative weights) the social network neighbors' average income, 3-clique, and income.

\paratitle{Generative process.}
We use Barabasi-Albert random graph generating algorithm \cite{cond-mat-0106096} to sample a random graph $(\set{V},\set{E})$ with $|\set{V}| = n$ nodes. Barabasi-Albert graphs have similar structure to several natural and human-made systems such as social networks, world wide web, citation networks etc \cite{dorogovtsev2003evolution}. 

For each unit $v_i \in \set{V}$, we have observed attributes such age, income, vaccine (a binary attribute indicating whether that unit was treated or not) etc. 
The probability of infection under no-vaccination, $infect_i(0)$, is a probabilistic function of individual's income, average income of her neighbors in the social network, membership in a 3-clique, and a mean zero Gaussian noise with variance $\sigma^2$. 
The probability of infection under vaccination, $infect_i(1)$, is constant at $0.05$. This can be thought of as the chance of getting infected at the vaccination center. Thus, the true effect modifiers are the income of the individual, average income of her neighbors in the social network and membership in a 3-clique. The propensity score for a unit's treatment is also a function of an individual's income, average income of their social network neighbors and their participation in 3-clique.

\paratitle{Testing for effect modifiers.}
Table \ref{tbl:synth_data} shows the results of running Algorithm \ref{algo:hypothesis_test} with all hypothesized covariates listed in the leftmost column for $4096$ units. The algorithm is able to find the true effect modifiers and gives the rest of the covariates an $\is$ value of $0$. 
When considering the $\ds$ values in the table, we see that these are not necessarily a good litmus test for true effect modifiers, since some of the covariates that are not effect modifiers got non-zero values, and it is not clear which threshold should be used to distinguish between these and the true effect modifiers. 

\begin{table}[]
    \caption{$\is$ and $\ds$ values using the synthetic infection data;  no. of samples $n=4096$,  noise variance  $\sigma^2=1$
    }
    \label{tbl:synth_data}
   \centering    
    \begin{tabular}{lrr}
    \toprule
        {Hypothesized Covariate} &         $\ds$ &         $\is$ \\
        \midrule
        Income          &  1.1657 &  14.23 \\
        Age          &  0.0003 &   0.00 \\
        Neighbors' Avg. Income &  5.3880 &  81.44 \\
        Neighbors' Avg. Age &  0.0020 &   0.00 \\
        Clique-3    &  5.1922 &  80.74 \\
    \bottomrule
    \end{tabular}

\end{table}

\begin{figure*}[t]
    \centering
    \begin{subfigure}{0.33\linewidth}
    \centering
    \includegraphics[width=\textwidth]{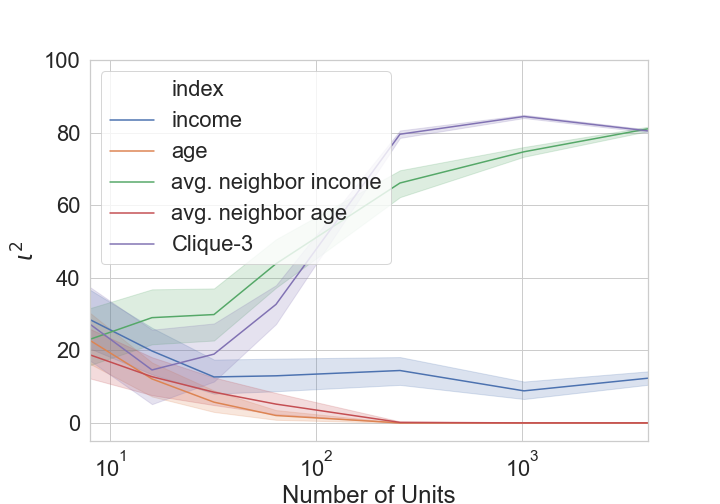}
    \caption{}
    \label{fig:num_units_a}
    \end{subfigure}%
    \begin{subfigure}{0.33\linewidth}
    \centering
    \includegraphics[width=\textwidth]{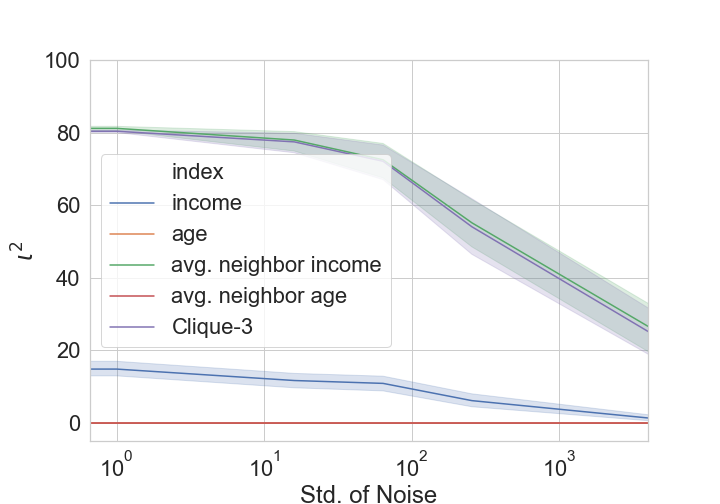}
    \caption{}
    \label{fig:noise}
    \end{subfigure}
       \begin{subfigure}{0.33\linewidth}
    \centering
    \includegraphics[width=\textwidth]{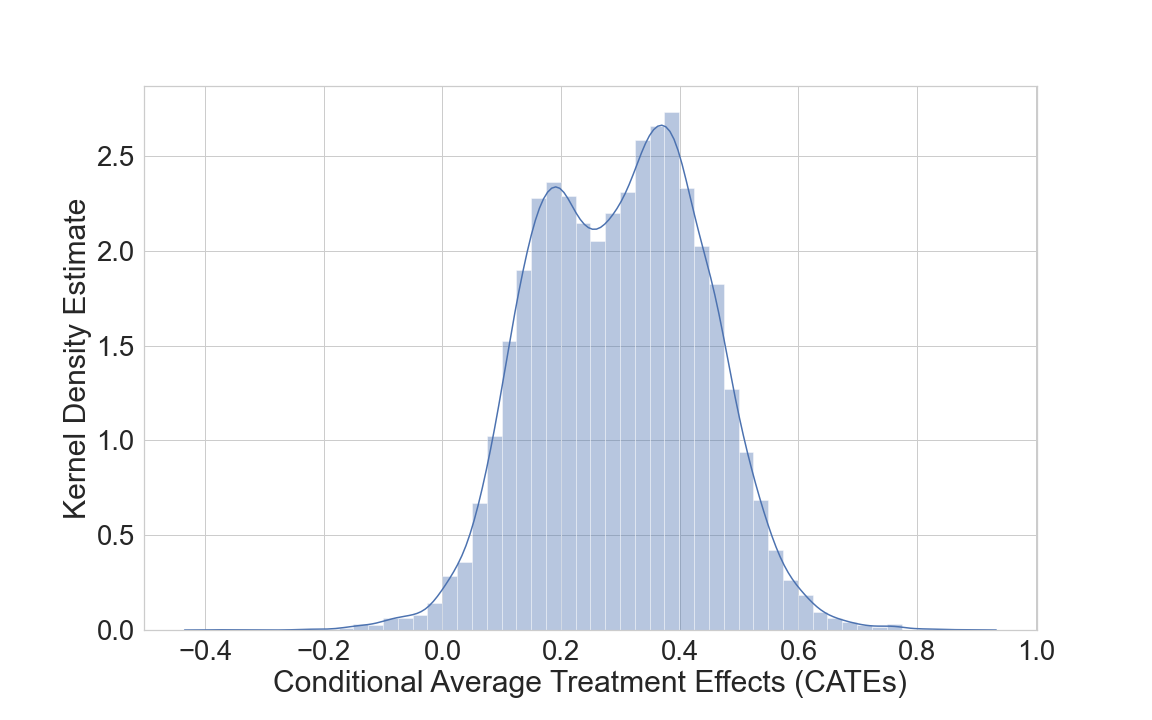}
    \caption{}
    \label{fig:density}
    \end{subfigure}%
    
        
        
    \caption{
    In (a, b), the green, purple and blue lines represent the true effect modifiers. (a) $\is$ as a function of the number of units in the synthetic data. (b) $\is$ as a function of noise in the potential outcome in the synthetic data. (Lines for avg. neighbor age and age are both at zero. (c) Kernel density estimation plot and histogram showing the distribution/frequency of CATE estimates across all units.)} \label{fig:num_units}
    \vspace{-3mm}
\end{figure*}

\paratitle{Effect of increasing the number of units on $\is$:}
We increased the number of units ($n$) from $8$ to $4098$ and computed the $\is$ values for each of the covariates with Algorithm \ref{algo:hypothesis_test}. 
As shown in Figure~\ref{fig:num_units}, for small values of $n$, the $\is$ value estimates have large variance.
However as $n$ grows larger than $200$ units, we observe that $\is$ estimates for the true effect modifiers (income of the individual, average income of her neighbors in the social network and membership in a 3-clique) are greater than $0$ and proportional to the size of their contribution to treatment effect (as per the data generative process mentioned above), i.e., membership in 3-clique and the effect of neighbors' average income are largest, followed by the effect due to one's own income.
Furthermore, we observed that as the number of units increased, the fluctuation in $\is$ decreased and stabilized to a value greater than zero for effect modifiers and zero for non effect modifiers (Figure \ref{fig:num_units}) 

\paratitle{Effect of increasing the noise levels on $\is$:}
In this part of the study we increased the variance of noise ($\sigma^2$) in the data generative process from $0$ to $2^{12}$, keeping the number of units constant at $n = 4000$. 
Figure~\ref{fig:noise} shows that the results of our algorithm are less sensitive to increase in the variance of noise. 
However, when the variance of noise is extremely large ($\sigma^2>2,500$), the $\is$ values of all covariates decrease, but maintain the same relationships. 
In particular, the $\is$ value for the average of neighbors' income has decreased from an average 80 to 30. 
We conclude that our hypothesis testing framework is not highly sensitive to noise in true data generative process, with inference based on $\is$ estimates being congruent to true data generative process for relatively large variance of noise in the outcome generative process.

\subsection{Real Data: Micro-finance/Risk Tolerance}\label{sec:real_data}
We analyze the causal effect of participating in self-help group (SHG) on financial risk tolerance using a $2010$ survey data from $77$ villages in Karnataka, India initially studied as part of \cite{banerjee2014gossip,jackson2012social}. 

\paratitle{Data properties and analysis:}
The survey data has $19$ features for $16,995$ individuals across $77$ villages including their age, occupation, gender, etc. Furthermore, the data also has $12$ different social networks of $69,000$ individuals (including $16,995$ surveyed individuals) across the same $77$ villages such friendships, relatives, social-visit networks, financial exchange etc. 
We consider all these connections in the same manner so that we have a homogeneous network, where all edges have the same interpretation. 
We use the individual's participation in SHG as the treatment indicator while using the indicator for an individual having a current outstanding loan or not, as the proxy indicator of their risk tolerance (i.e., the outcome).

\begin{figure}[t]
    \centering
    \includegraphics[width=0.42\textwidth]{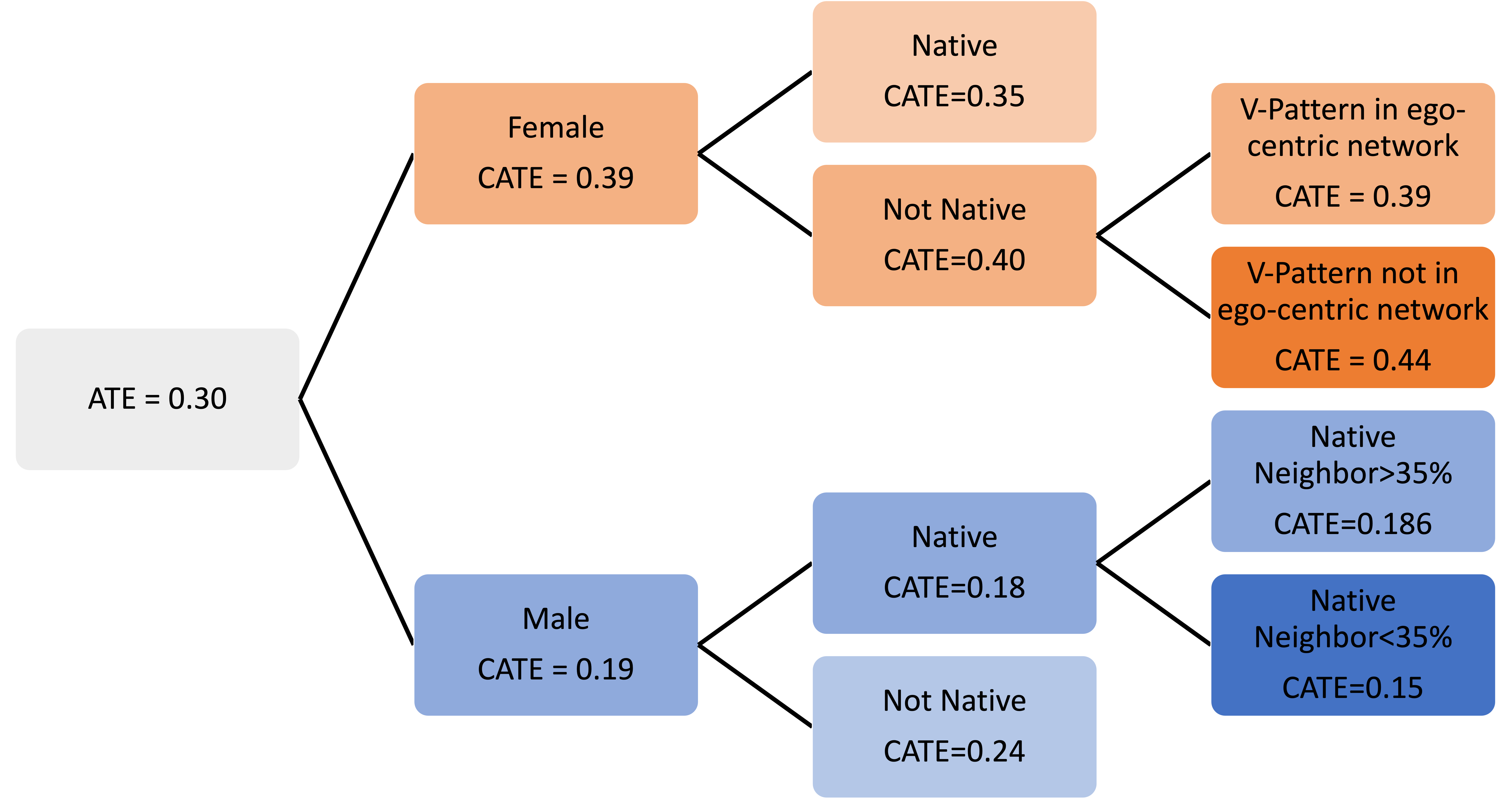}
    \caption{
    Recursive partitioning of the covariate space and corresponding CATEs}
    \label{fig:tree}
    \vspace{-3mm}
\end{figure}

Some units have relatively low treatment effect between 0.0 and 0.2, while the other units have higher treatment effect ranging between 0.4 to 0.6. The heterogeneity across units is evident from the kernel density estimate of CATEs (shown  in Figure~\ref{fig:density}).


\begin{table}[]
\centering
\caption{
$\is$ and $\ds$ values for CATE and CATT estimates using microfinance data \cite{banerjee2011poor,banerjee2014gossip}
}
\label{tbl:case_study}
\begin{tabular}{l|rr|rr}
\hline
                       & \multicolumn{2}{c|}{CATE} & \multicolumn{2}{c}{CATT}                                  \\ \hline
Hypothesized Covariate & $\ds$     & $\is$     & \multicolumn{1}{l}{$\ds$} & \multicolumn{1}{l}{$\is$} \\ \hline
Age                    & 0.21        & 0.00        & 0.27                        & 0.00                        \\
Education              & 0.57        & 0.00        & 0.089                       & 0.00                        \\
Native                 & 2.66        & 0.62        & 0.03                        & 0.00                        \\
Female                 & 5.94        & 0.83        & 0.02                        & 0.00                        \\
\% neighbor Native     & 1.99        & 0.49        & 0.19                        & 0.00                        \\
\% neighbors Female    & 3.35        & 0.70        & 0.40                        & 0.00                        \\
Avg. neighbor's Age    & 0.64        & 0.00        & 3.28                        & 0.69                        \\
Clique-3               & 0.026       & 0.00        & 0.511                       & 0.00                        \\
V-Pattern              & 0.020       & 0.00        & 4.07                        & 0.75                        \\ \hline
\end{tabular}
\end{table}

\paratitle{Testing for effect modifiers:}
We assume that all pre-treatment covariates of an individual or the individual's social connection are potential effect modifiers. Further, using the pattern mining algorithm we further hypothesized that being part of network structures like 3-cliques or `V's can be potential effect modifiers. 
For our study, the estimands of interest are CATE and CATT for different levels of $\set{X}$, $\omega$ and $\eta$. As mentioned in Section~\ref{sec:single}, we used non-parametric double machine learning to estimate CATE and CATT. 

We test for effect modifier hypotheses using Algorithm \ref{algo:hypothesis_test} with gradient boosting regression for posterior projection (line \ref{l:project}). Table \ref{tbl:case_study} shows the $\is$ values for hypothesized CATE and CATT effect modifiers. 
We report the test statistics for the open and closed triadic financial cooperation, age, gender, nativeness, education, some of neigbhors' covariates etc. 
We observe that having gender, nativeness, the percentage of female neighbors, and the percentage of native neighbors are all CATE effect modifiers for CATE. However, when we test for the CATT effect modifiers, we find that average age of the neighbors and V-pattern in ego-centric subgraph (open triad) are effect modifier while the test fails to reject the null for other variables. 

\paratitle{Societal Impact and Implications.}
Based on our results in Table~\ref{tbl:case_study} and Figure~\ref{fig:tree}, we conjecture that SHG are highly beneficial to females or individuals who are non-native to the village. The benefits of SHG are amplified if the pretreatment network of the individual is weak. Typically, these individuals might be poorer and might find it hard to borrow money. However, with the help of SHG such individuals can borrow money at a reasonable rate where the pooled capital of SHG helps them mitigate the risk of a default. Thus, if a government or a policy maker were to promote SHG for increasing financial risk tolerance and entrepreneurship in financially weaker section of society, it might be optimal to engage with females and migrants. 
Further, targeting 
individuals at the periphery of social network is also important. 
However, while studying treatment effect modifiers, one must be careful about potential misuses and fairness, e.g., even the sub-population who do not benefit the most from a vaccine should be able to access a vaccine. 

\section{Related work}\label{sec:related}\vspace{-2mm}
There is rich literature on {\bf statistical tests for hypothesis testing} 
(e.g., \cite{snyder1978hypothesis,shaffer1995multiple,list2019multiple,xia2017hypothesis,LAROSA201581}). %
Snyder et. al. \cite{snyder1978hypothesis} proposes an approach for allowing people to test hypotheses about other people through their social interactions.
Multiple hypotheses testing \cite{shaffer1995multiple,list2019multiple} focuses on examining multiple hypotheses simultaneously and includes a consideration of their possible interactions with each other, as well as their closure etc. 
One common hypothesis is that of heterogeneity in different studies. 
Therefore, many previous works have devised methods for heterogeneity detection (e.g., \cite{hardy1998detecting,crump2008nonparametric,green2012modeling,taddy2016nonparametric}). These works develop and evaluate tests that quantify the amount of heterogeneity in experiments w.r.t. the number of experiments included, the total information available, and the distribution of weights among the different experiments. 
As mentioned in the paper, our test is analogous to the $I^2$ test for heterogeneity  \cite{higgins2002quantifying,higgins2003measuring}.
%
{\bf Effect modification} has also been explored in different contexts \cite{katsouyanni2001confounding,greenland1989ecological,knol2012recommendations,vanderweele2009distinction,vanderweele2007four,hernan2010causal,blot1979synergism,rothman1980concepts,saracci1980interaction}. %
Works such as \cite{blot1979synergism,rothman1980concepts,saracci1980interaction} have pointed at the relevance of effect modification to public health, while others \cite{vanderweele2009distinction,knol2012recommendations} focused on the analysis of effect modifiers. 
In general, these works do not propose a testing framework for effect modifiers that is specifically adapted to social networks. 
%
Multiple previous works has studied {\bf causal inference in the presence of social network}
\cite{graham2010measuring,halloran2012causal,halloran1995causal,vanderweele2011bounding,aronow2017,shalizi2011homophily,ogburn2017causal}. 
These works address applications such as the study of infectious diseases \cite{vanderweele2011bounding, halloran1995causal} or behavior and, in particular, interactions in social networks \cite{shalizi2011homophily,ogburn2017causal, sobel2006randomized, vanderweele2011bounding,hong2006evaluating}. 
Our work, on the other hand, focuses on developing a framework for testing for causal effect modifiers in a network. 
{\bf Social network analysis} (e.g., \cite{sailer1978structural,knoke2019social,wasserman1994social}) focuses on gaining insights from social networks using graph theory and graph mining. One considered aspect is the structure of the network and the local neighborhoods of the actors (nodes) in the network. Some structures, such as centrality \cite{valente2008correlated,gest2001peer} and triangles \cite{granovetter1973strength,wasserman1994social} are of particular interest as they have been correlated with other semantic traits of the actors in them. In addition, there has been work on detecting the equivalence of actors based on their ego-centric graph  \cite{everett1990ego,pattison1993algebraic}. 


\vspace{-2mm}\section{Conclusions}\label{sec:conclusions}\vspace{-2mm}
We have devised and studied a novel test for effect modifiers in social networks. 
We have provided desiderata and algorithms for obtaining effect modifiers based on the network structure and for testing whether a covariate is an effect modifier. 
Our experimental evaluation suggests that our test and framework are able to detect effect modifiers and avoid false positives, while the use-case we include revealed effect modifiers related to the environment of the units in the social network. 
Intriguing directions of future work include the extension of our framework to non-homogeneous social networks where different units and different ties can be interpreted in a different manner (e.g., edges with different labels), 
and improving the efficiency of our pattern mining procedure.

\clearpage
\balance
\bibliographystyle{acm}
\bibliography{biblio}

\clearpage
\appendix
\section{Appendix}\label{sec:appendix}
We next give the proofs of the theorems from the paper, describe our procedure for mining network patterns that are suspected effect modifiers, detail how our Algorithm \ref{algo:hypothesis_test} can test covariates and sets of effect modifier, and give more details about the experiments and implementation.

\subsection{Proofs of the Theorems}

\begin{proof}[Proof of Theorem \ref{thm:identify}] 
Using the structural equation for $Y$, we know that potential outcome 
\begin{equation*}
    Y_i(t) = \phi_Y\left(\set{X}_i, \{\set{X}_j\}_{v_j\in \set{V}_{\set{G}_i}\setminus\{v_i\}}, \set{E}_{\set{G}_i}, t \right) + \epsilon^Y_i.
\end{equation*} 
Thus, if $T_i=t$ then $Y_i(t) = Y_i$ by definition.
Without loss of generality, let's assume $T_i=1$, then 
\begin{equation*}
    E[Y_i(1) | \set{X}_i, \{\set{X}_j\}_{v_j\in \set{V}_{\set{G}_i}\setminus\{v_i\}}, \set{E}_{\set{G}_i}] = E[Y_i | \set{X}_i, \{\set{X}_j\}_{v_j\in \set{V}_{\set{G}_i}\setminus\{v_i\}}, \set{E}_{\set{G}_i}, T_i=1].
\end{equation*}
By assumption A.4 (``Distributional assumptions'' in Section \ref{sec:pcmsn}). 
\begin{equation*}
    E[Y_i | \set{X}_i, \{\set{X}_j\}_{v_j\in \set{V}_{\set{G}_i}\setminus\{v_i\}}, \set{E}_{\set{G}_i}, T_i=1] = \phi_Y(X_i,\{X_j\}_{j\in N_i\setminus\{i\}}, EN_i, T_i).
\end{equation*}
Now, by the positivity assumption (Section \ref{sec:pcmsn}), $\exists k$ such that $T_k = 1-T_i$, $\set{X}_i=\set{X}_k$, $\omega_i=\omega_k$ and $\eta_i=\eta_k$.\\ Further, by summarizability assumptions (Section~\ref{eq:assumption_summary}) and A.4. 
\begin{equation*}
E[Y_i(0) | \set{X}_i, \{\set{X}_j\}_{v_j\in \set{V}_{\set{G}_i}\setminus\{v_i\}}, \set{E}_{\set{G}_i}]=E[Y_i(0) |\set{X}_i, w_i, \eta_i] = \phi_Y(\set{X}_i, w_i, \eta_i, 0)
\end{equation*}
Thus, following the argument,
$$\phi_Y(\set{X}_i, w_i, \eta_i, 0) =
\phi_Y(\set{X}_k, w_k, \eta_k, 0)$$
By assumption A.4.
\begin{equation*}
\phi_Y(\set{X}_k, w_k, \eta_k, 0) = E[Y_k(0) | X_k,w_k, \eta_k]=E[Y_k| X_k,w_k, \eta_k,T_k=0]
\end{equation*}
Hence,
\begin{equation*}
    \tau_i = E[Y_i| \set{X}_i,w_i, \eta_i,T_i=1]-E[Y_k| \set{X}_k,w_k, \eta_k,T_k=0]
\end{equation*}
such that $Z_k = 1-Z_i$, $X_i=X_k$, $w_i=w_k$ and $\eta_i=\eta_k$.
The above proof proves that the conditional average treatment effect of interest is identifiable in terms of the observables.
\end{proof}

\begin{proof}[Proof of Theorem \ref{thm:consistency}]
As the causal effect estimator is consistent, we know that $\widehat{\bar{\tau}} \to \bar{\tau}$ and $\widehat{\tau}(w) \to \tau(w)$ when $n\to\infty$. If feature $W$ is not an effect modifier then $(Y(1) - Y(0)) \perp W$. Hence, $E[Y(1) - Y(0)| W=w] = E[Y(1) - Y(0)]$ for all $W=w$. Thus, $\widehat{\ds}_W \to 0$ as $n \to \infty$. This further implies $\widehat{\is}_W\to 0$ as $n \to \infty$. Hence, the test statistic is consistent under null.
\end{proof}

\subsection{Algorithm for Mining Network Patterns in Section \ref{sec:single}}\label{sec:appendix-graph}

We next describe our greedy algorithm for finding a pattern suspected as being an effect modifier. After finding such a pattern, our framework can be employed to verify whether this pattern is indeed an effect modifier. 
Since the objective function in Section \ref{sec:single} aims to find the pattern $\Delta$ that maximizes an expression over all patterns, a na\"ive algorithm would iterate over all possible patterns in the network $\set G$, which would require iterating over an exponential number of patterns (exponential in the size of $\set G$).

Instead, we propose a greedy approach that eliminates the need to iterate over all patterns. 
Intuitively, the algorithm attempts to find the largest pattern that is found in the ego-centric networks of the nodes with the highest outcomes, while still satisfying the condition in the optimization problem. 
Algorithm \ref{algo:pattern} gets as input the social network, a list of pairs of nodes and their respective outcomes, the parameter $\delta$ that determines the outlier sensitivity, and a threshold for minimal pattern size $c$. 
It first sorts the nodes by their outcome, sets the initial pattern to be the ego-centric network of the node with the highest outcome, and initializes an index $j$ (lines \ref{l:sort}--\ref{l:index}). 
Next, the algorithm updates the pattern in a while loop that runs as long as the condition in the optimization problem is not satisfied (line \ref{l:while1}). 
It does so by finding the largest common subgraph between the current pattern and the ego-centric network of the next node in the sorted list (function $LCS$ in line \ref{l:lcs1}), and if the updated pattern is larger than the specified threshold, it is updated (lines \ref{l:threshold}--\ref{l:update1}). 
The second while loop considers more nodes from the sorted list and tries to increase the objective expression while also ensuring that the condition in the optimization function is still satisfied (lines \ref{l:while2}--\ref{l:index2}). Finally, it returns the generated pattern in line \ref{l:return-pattern}.

\begin{small}
\IncMargin{1.1em}
\begin{algorithm}[!htb]
	\SetKwFunction{Test hypothesized effect modifier}{Test hypothesized effect modifier}
	\SetKwInOut{Input}{input}\SetKwInOut{Output}{output}
	\LinesNumbered
	\Input{Social network $\set{G}$, a list $L_{Y}$ of pairs $(i, Y_i)$ where $i$ is a node and $Y_i$ is the observed outcome, outlier sensitivity $\delta$, minimal pattern size $c$}
    \Output{Candidate pattern $\Delta^*$}
	\BlankLine
    
    $S_{Y} \gets sort(L_{Y}, by=Y_i,\text{descending})$\;\label{l:sort}
    $\Delta^* \gets \set{G}_{S_{Y}[0][0]}$\;
    $j \gets 1$\;\label{l:index}
    \While{$\left|\frac{\sum_i \mathds{1}[\Delta^* \in \set{G}_i]}{n} - \frac{\sum_i \mathds{1}[\Delta^* \notin \set{G}_i]}{n}  \right| > \delta$}
    {\label{l:while1}
        $\Delta^{'} \gets LCS(\Delta^*,\set{G}_{S_{Y}[j][0]})$\;\label{l:lcs1}
        \If{$|\Delta^{'}| > c$}
        {\label{l:threshold}
            $\Delta^* \gets \Delta^{'}$\;\label{l:update1}
        }
        $j \gets j+1$
    }
    \While{$\left|\frac{\sum_i \mathds{1}[\Delta^* \in \set{G}_i]}{n} - \frac{\sum_i \mathds{1}[\Delta^* \notin \set{G}_i]}{n}  \right| \leq \delta$}
    {\label{l:while2}
        $\Delta^{'} \gets LCS(\Delta^*,\set{G}_{S_{Y}[j][0]})$\;\label{l:lcs2}
        \If{$Objective(\Delta^{'})>Objective(\Delta^{^*})$ and $|\Delta^{'}| > c$}
        {
            $\Delta^* \gets \Delta^{'}$\;\label{l:update2}
        }
        $j \gets j+1$\;\label{l:index2}
    }
    \Return $\Delta^*$\;\label{l:return-pattern}
	\caption{Find suspected pattern}
	\label{algo:pattern}
\end{algorithm}
\end{small}

Checking whether the ego-centric network of a node contains a given pattern in lines \ref{l:while1}, \ref{l:while2} and finding the largest common subgraph in two graphs in lines \ref{l:lcs1}, \ref{l:lcs2} incurs exponential complexity in the size of the ego-centric networks as this is equivalent to checking subgraph isomorphism. 
Previous work on this subject has proposed approximation methods and heuristics \cite{CHEN2015265}.
In our implementation, we have used the ISMAGS algorithm \cite{houbraken2014index}, but other approaches can certainly be plugged in to our algorithm as black-boxes. 

We further employ an optimization that reduces the number of subgraph isomorphisms that need to be considered in each iteration of the two while loops (lines \ref{l:while1} and \ref{l:while2}). For nodes that have been considered in previous iterations in the list $L_{Y}$ and whose pattern $\Delta^*$ was updated in lines \ref{l:update1}, \ref{l:update2}, we do not need to repeatedly check whether their ego-centric network contains $\Delta^*$, since we know that any subsequent pattern will be a subgraph of $\Delta^*$ and therefore, their ego-centric network will contain it as well.  

\subsection{Testing Different Forms of Hypothesized Effect Modifiers}
Our framework supports the testing of network patterns, covariates, summarized neighbor covariates, and sets of hypothesized effect modifiers of different types, as mentioned in Section \ref{sec:criteria_test}. 
Using our notation, the set $\set{W}$ can contain different combinations of hypothesized effect modifiers. 

If $\set{W}$ contains a covariate $X \in \set{X}_i$ of a unit $v_i$ (option (1) in Section \ref{sec:criteria_test}), Algorithm \ref{algo:hypothesis_test} remains unchanged and works in the same manner as for a network pattern, i.e., by computing $\hat{\is}_W$ and checking whether $\hat{\is}_W > I_0$ for a threshold $I_0$. 
$\set{W}$ can also contain a summarized covariate of the neighbors of a unit (option (2) in Section \ref{sec:criteria_test}). 
Such summarized covariates can be obtained by using a summary function over the covariates of the neighbors of each node, creating a single scalar that represents a specific covariate of all neighbors, regardless of their number. We use mean, but different summary functions can also be employed. 
If $\set{W}$ contains a summarized covariate of the neighbors, Algorithm  \ref{algo:hypothesis_test} again operates in the same manner since there is a single summarized covariate for each node. 
If $\set{W}$ contains a set of hypothesized effect modifiers, Algorithm \ref{algo:hypothesis_test} is applied to each of the elements of $\set{W}$ separately to test if they are marginal effect modifiers. In this paper, we do not check if all possible subsets of $\set{W}$ are effect modifiers jointly, however, this framework trivially generalizes to that case.

\subsection{Generative Process for Synthetic Data in Section \ref{sec:synthetic}}
We use Barabasi-Albert random graph generating algorithm \cite{cond-mat-0106096} to sample a random graph $(\set{V},\set{E})$ with $|\set{V}| = n$ nodes. Barabasi-Albert graphs have similar structure to several natural and human-made systems such as social networks, world wide web, citation networks etc \cite{dorogovtsev2003evolution}. 
For each unit $i \in \set{V}$, we define three attribute: age, income and vaccine (a binary attribute indicating whether that unit was vaccinated or not). 
We use the following data generative process: 
$age_i \sim Uniform(21,99)$, $income_i \sim Uniform(20,60) + Normal\left(\frac{age_i}{50},5^2 \right)$, $vaccine_i \sim Bernoulli(1/2)$. 
The potential outcome under placebo, $infect_i(0)$, is a function of individual's income, average income of her neighbors in the social network, membership in a 3-clique, and a mean zero Gaussian noise with variance $\sigma^2$. 
The potential outcome under vaccination, $infect_i(1)$, is constant and equals $0.1$. Thus, the true effect modifiers are the income of the individual, average income of her neighbors in the social network and membership in a 3-clique. 
We define $\epsilon_i \sim \mathcal{N}(0,\sigma^2)$, 
\begin{footnotesize}
\begin{equation*}
    p_i = expit\left( 200-income_i -
    \frac{4\sum_{j\in\set{V}_{\set{G}_i}} income_{j}}{|\set{V}_{\set{G}_i}|} +  4triangle_i + \epsilon_i 
    \right),
\end{equation*}
\end{footnotesize}
$infect_i(0) = Bernoulli(p_i)$, and $infect_i(1) = Bernoulli(\frac{1}{10})$.
We generated the synthetic data keeping real world dynamics in mind, i.e., higher income individuals can potentially work from home which will reduce their chances of being infected by SARS-Cov-2 virus. Similarly, if the social connections of an individual have high income, then it will reduce their chances of exposure to the virus. Lastly, if an individual is not part of a 3-clique then they are less likely to meet an individual, which will further reduce the chances of infection. We assume that post-infection, every individual in the population has an equal chance of getting infected which is a small non-zero probability.

\end{document}